\documentclass[%
 reprint,
superscriptaddress,
 amsmath,amssymb,
 aps,
prl,
showkeys
]{revtex4-2}

\usepackage{graphicx}
\usepackage{dcolumn}
\usepackage{bm}
\usepackage{color}

\newcommand{\probP}{\text{I\kern-0.15em P}}

\begin{document}

\preprint{APS/123-QED}

\title{Langevin equation in heterogeneous landscapes: how to choose the interpretation}

\author{Adrian Pacheco-Pozo}
    \affiliation{Department of Electrical and Computer Engineering and School of Biomedical Engineering, Colorado State University, Fort Collins, CO 80523, USA}
 
\author{Micha{\l} Balcerek}%
    \affiliation{Faculty of Pure and Applied Mathematics, Hugo Steinhaus Center, Wroc{\l}aw University of Science and Technology, 50-370 Wrocław, Poland}%

\author{Agnieszka Wy{\l}omanska}
    \affiliation{Faculty of Pure and Applied Mathematics, Hugo Steinhaus Center, Wroc{\l}aw University of Science and Technology, 50-370 Wrocław, Poland}%

\author{Krzysztof Burnecki}
    \affiliation{Faculty of Pure and Applied Mathematics, Hugo Steinhaus Center, Wroc{\l}aw University of Science and Technology, 50-370 Wrocław, Poland}%

\author{Igor M. Sokolov}
\affiliation{Institut f\"{u}r Physik and IRIS Adlershof, Humboldt University Berlin, Newtonstrasse 15, 12489 Berlin, Germany}

\author{Diego Krapf}
    \email{Corresponding author: diego.krapf@colostate.edu}
    \affiliation{Department of Electrical and Computer Engineering and School of Biomedical Engineering, Colorado State University, Fort Collins, CO 80523, USA}

\date{\today}

\begin{abstract}
The Langevin equation is a common tool to model diffusion at a single-particle level. In non-homogeneous environments, 
such as aqueous two-phase systems or biological condensates with different diffusion coefficients in different phases, 
the solution to a Langevin equation is not unique unless the interpretation of stochastic integrals involved is selected. 
We analyze the diffusion of particles in such systems and evaluate the mean, the mean square displacement, and the 
distribution of particles, as well as the variance of the time-averaged mean-squared displacements. Our analytical 
results provide a method to choose the interpretation parameter from single particle tracking experiments.
\end{abstract}

\maketitle


Diffusion is a fundamental process that is ubiquitous in physics, chemistry, and biology. The first successful attempts to understand such motion came from Einstein and, independently, from Smoluchowski, who described the probabilistic nature of Brownian motion \cite{Einstein1905,Smoluchowski1906}. Shortly after these seminal works, Langevin introduced a single-particle formalism equivalent to Newton's second law of motion in the context of statistical physics \cite{Langevin1908}.
The Langevin equation is a differential equation for the position $X(t)$ of a tracer particle that includes a random force $\xi(t)$ related to the interactions of the tracer with the particles forming the medium  \cite{Karatzas2012,Oksendal2013,Braumann2019}. The strength of the random force is proportional to the square root of the diffusion coefficient $D$. Thus, a heterogeneous environment involving variations in local diffusivity yields local fluctuations in the random forces.

During the last decade, heterogeneous processes have received increased attention due to the broad use of single particle tracking methods and their sensitivity to probe local diffusive properties \cite{vanMilligen2005,Cherstvy2013,Manzo2015,Weron2017,Tsekouras2015,shen2017single,norregaard2017manipulation,Chechkin2017,krapf2019strange,Postnikov2020,Janczura2021,Pacheco2021,Singh2022,munoz2023quantitative,Pacheco2023,korabel2010paradoxes}.
In particular, systems having a boundary that separates two liquid phases where particles have different diffusivities are gaining technological interest in food science, chemical synthesis, and biomedical engineering \cite{munchow2008protein,chao2020emerging,keating2021liquid}. Aqueous two-phase systems are spontaneously formed by the separation of two incompatible polymers above a critical concentration. Their key property is that small molecules can diffuse across the interface. Currently, they are employed in the separation and purification of biomolecules \cite{albertsson2020aqueous}, carbon nanotubes \cite{yang2022polyoxometalate}, and metal ions \cite{rahimipoor2023investigation}. Furthermore, liquid phase separation has been recently found to play fundamental roles in cellular processes, such as genetic regulation \cite{henninger2021rna} and signaling cascades \cite{lyon2021framework}. In the life sciences, these compartments are thought of as membraneless organelles composed of dense assemblies of proteins and nucleic acids and are termed biological condensates \cite{bo2021stochastic,hubatsch2021quantitative}. Thus, modeling the diffusion of molecules across these boundaries has importance both from scientific and technological perspectives. In addition to two-phase systems, heterogeneous diffusion has been probed in a wide variety of systems, including porous media \cite{Berkowitz2002}, supercooled liquids \cite{Yamamoto1998}, and micropillar matrices \cite{Chakraborty2020}. 

The Langevin equation for a particle in a homogeneous environment is well-defined \cite{Sokolov2010}. However, for heterogeneous environments, the integrals appearing when solving the Langevin equation are not uniquely defined. The lack of uniqueness is rooted in the integrand function having a random nature. The problem becomes severe when the diffusivity landscape has abrupt changes, such as those encountered in a two-phase system. In practice, stochastic integrals are defined as sums over infinitesimal rectangles, and, contrary to the case of smooth functions, the integral depends on the
position of the points where the function is evaluated. 
This dependence, thus, requires additional information known as interpretation \cite{deHaan2012,Escudero2023,vanKampen1981,Sancho2011,Tsekov1997,Farago2014,menon2023random}. Due to its lack of a unique solution, some authors prefer to term the Langevin equation a pre-equation \cite{vanKampen1981}.

The first approach to solving the Langevin equation via stochastic integration was introduced by It\^o, employing the initial point of the infinitesimal interval to compute the integrals \cite{Ito1944}. Then, Stratonovich introduced an alternative method by which the midpoint of the interval is selected \cite{Stratonovich1964,Stratonovich1966}. Finally, H\"anggi and Klimontovich introduced yet another prescription, taking the last point of the infinitesimal interval \cite{Hanggi1978,Hanggi1980,Hanggi1982,Klimontovich1994}. Such freedom at selecting an interpretation gives rise to the problem of choosing the \textit{`correct'} interpretation \cite{west1979stochastic}. 
Thus, when developing models to describe dynamics in heterogeneous environments a choice is made \textit{a priori}, typically employing either It\^o \cite{cumberworth2023constriction}, Stratonovich \cite{Cherstvy2013}, or H\"anggi-Klimontovich (HK) \cite{cherstvy2014particle} prescriptions.   

Nowadays, it is accepted that the choice of interpretation depends on the underlying mechanisms that dictate the fluctuations of the physical system \cite{vanKampen1981,Sokolov2010}. However, methods that assign an interpretation of Langevin equation to experimental physical systems are still missing. Given the broad interest in diffusion in heterogeneous media, this problem and its implications need careful consideration. Particularly, tools to infer the interpretation parameter from physical observables can help guide experimentalists and theoreticians in the use of Langevin equation.
 
In this letter, we study the diffusion of particles in a heterogeneous environment that mirrors a two-phase system, namely, having different diffusion coefficients on each side of an interface. We consider the interpretation as a parameter, focusing on the It\^o, Stratonovich, and HK interpretations. 
The simple situation considered here 
elucidates significant variations between solutions associated with different interpretations and leads to distinctive experimental predictions.
We observe that typical characterizations, such as the mean, the mean square displacement (MSD), and the displacement probability density function (PDF) for different interpretations have different forms. Our results provide a tool to infer the interpretation parameter in experimental settings from the measured density of particles at both sides of an interface.

We start by considering a Langevin equation with position-dependent diffusion coefficient $D(x)$ \cite{dePirey2022},
\begin{equation}
d X(t) = \sqrt{2D[X(t)]} \; dB(t),
\label{eq:langevin_eq}
\end{equation}
where $B(t)$ is standard Brownian motion. In the mathematical literature, such a process is said to have \textit{multiplicative noise}, as opposed to \textit{additive noise} where the diffusion coefficient is either constant or a function of time \cite{dePirey2022}. To integrate this equation, one splits the time interval $[0,t]$ into $N$ subintervals of size $\Delta t = t/N$,
leading to $N$ integrals that are Riemann approximated,
\begin{equation}
\int_{t_n}^{t_{n+1}} \sqrt{2D[X(s)]} \; dB(s) \\
\approx \sqrt{2D[X(t^{\prime})]} \; \xi_n,
\label{eq:Riemann} 
\end{equation}
where ${t_n = n \Delta t}$, $n = 0, \dots, N$, $t^{\prime} \in [t_n,t_{n+1}]$, and $\xi_n = B(t_{n+1})-B(t_n)$ are the increments of Brownian motion, i.e., independent and identically distributed Gaussian random variables with variance $\Delta t$. In contrast to additive noise, the choice of $t^{\prime}$ alters the evolution of the system because  
different times within $[t_n,t_{n+1}]$ can give different diffusion coefficients. 

Let us express $t^{\prime}$ as
$t^{\prime} = t_n + \alpha ( t_{n+1} - t_n )$,
with $\alpha \in [0,1]$ known as the interpretation parameter. Performing a Taylor expansion to first order of $X(t^{\prime})$ around $t_n$, we find $X(t^{\prime}) \approx X(t_n) + \alpha \Delta X_n$, with $\Delta X_n = X(t_{n+1}) - X(t_n)$, and by replacing this expression in Eq.~(\ref{eq:Riemann}), 
\begin{equation}
X(t) = \sum_{n=0}^{N-1} \sqrt{ 2 D\left[X(t_n) + \alpha \Delta X_n\right]} \; \xi_n.
\end{equation}

The Langevin equation with multiplicative noise (Eq.~(\ref{eq:langevin_eq})) is associated with a Fokker-Planck equation for the PDF $p(x,t)$ of the particle's position at time $t$ \cite{Sokolov2010,dePirey2022} 
\begin{equation}
\frac{\partial p(x,t)}{\partial t} = \frac{\partial}{\partial x}  \left\{ D^{\alpha}(x) \frac{\partial}{\partial x} \left[ D^{1-\alpha}(x) p(x,t) \right] \right\},
\label{eq:Fokker_Planck}
\end{equation}
where $\alpha$ is the interpretation parameter.

To numerically study this Fokker-Planck equation, we consider a discrete space with lattice constant $a$. 
Eq.~(\ref{eq:Fokker_Planck}) becomes
\begin{multline}
\frac{d p_i(t)}{d t} = \frac{ D^{\alpha}_{i-1/2} D^{1-\alpha}_{i-1}}{a^2} \; p_{i-1}(t) -  \\
- \frac{D^{\alpha}_{i+1/2} D^{1-\alpha}_{i} + D^{\alpha}_{i-1/2} D^{1-\alpha}_{i} }{a^2} \; p_{i}(t) + \\
+ \frac{ D^{\alpha}_{i+1/2} D^{1-\alpha}_{i+1}}{a^2} \; p_{i+1}(t),
\label{eq:master} 
\end{multline}
where $p_i(t)$ is the probability to find the particle on site $i$ at time $t$, that is $p_i(t)=p(x=ia,t)$, and $D_{i\pm 1/2}=D[x=(i\pm 1/2)a]$.
This equation has the form of a master equation \cite{Klafter2011} with transition rates $\omega_{i \to j}$ from position $i$ to $j$,
\begin{equation}
\omega_{i \to j} = \frac{ D^{1-\alpha}_i D^{\alpha}_{k} }{a^2},
\label{eq:trans_rates}
\end{equation}
with $k = (i + j)/2$ for $j = i \pm 1$.

\begin{figure*}
    \centering
    \includegraphics[width=\textwidth]{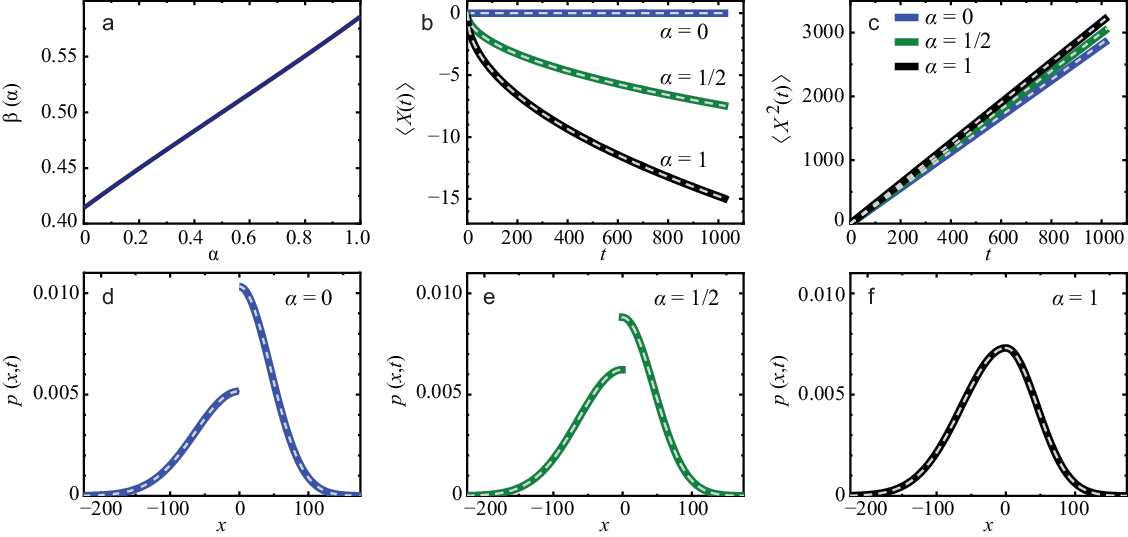}
    \caption{ Characterization of heterogeneous Brownian motion with $D_-=2$ and $D_+=1$.
    (a) Probability $\beta$ of finding the tracer in the left half plane as a function of $\alpha$, as found analytically (Eq.~(\ref{eq:beta})).
    (b) Mean $\langle X(t) \rangle$ for three interpretation parameters $\alpha = 0, 1/2,$ and $1$, corresponding, respectively to It{\^o}, Stratonovich, and H{\"a}nggi-Klimontovich. Thick lines are numerical simulations and thin dashed lines are analytical solutions.
    (c) MSD $\langle X^2(t) \rangle$ for the interpretation parameter $\alpha = 0, 1/2,$ and $1$. Thick lines are simulations and thin dashed lines are analytical solutions.
    (d-f) PDF at time $t = 1024$ for the three interpretations. Thick lines are numerical simulations of the master equation (Eq.~(\ref{eq:master})) and thin dashed lines are analytical solutions given by Eq.~(\ref{eq:gen_PDF}).
    }
    \label{fig:PDF}
\end{figure*}

We consider the master equation using a simple diffusion coefficient function of the form
\begin{equation}
D(x)= \left\{ 
\begin{array}{lcc} 
D_- & \text{if} & x < 0, \\ \\
D_+ & \text{if} & x \geq 0,
\end{array} 
\right.
\end{equation}
with $D_{\pm} > 0$. This function represents a two-phase system with a diffusivity change at the boundary. Setting $a = 1$, we use a fourth-order Runge-Kutta integration scheme \cite{NumericalRecipes} with an integration step of $\Delta t = 0.01$ to numerically solve the above master equation for different values of the interpretation parameter $\alpha$. We found that the PDF for any $\alpha$ is described by a \textit{generalized} two-piece Gaussian distribution \cite{SupplMat}  
\begin{equation}
p(x,t; \alpha)= \left\{ 
\begin{array}{lc} 
\displaystyle \frac{2 \beta(\alpha)}{\sqrt{4 \pi D_- t }} \exp \left(- \frac{x^2}{4 D_- t} \right), & x < 0, \\ \\
\displaystyle \frac{2 [1 - \beta(\alpha)]}{\sqrt{4 \pi D_+ t }} \exp \left(- \frac{x^2}{4 D_+ t} \right), & x \geq 0,
\end{array} 
\right.
\label{eq:gen_PDF}
\end{equation}
where $\beta(\alpha)$ is the probability that the particle is in the negative part of the axis, i.e. $\beta=\probP[X(t)<0]$, which is independent of time.
We verified that the identified PDF (Eq.~(\ref{eq:gen_PDF})) is the solution to the Fokker-Planck equation (Eq.~(\ref{eq:Fokker_Planck})) \cite{SupplMat}.
Then, using matching conditions on both sides of the origin, we find the probability $\beta$ \cite{SupplMat},
\begin{equation}
 \beta(\alpha) = \left[ 1 + \left( \frac{D_-}{D_+} \right)^{(1/2-\alpha)} \right]^{-1}.
\label{eq:beta}
\end{equation}

We focus on the first two moments, corresponding to the mean position and the MSD,
\begin{equation}
\langle X(t) \rangle = \frac{2}{\sqrt{\pi}}  \left[ \sqrt{D_+} - \beta(\alpha) \left( \sqrt{D_-} + \sqrt{D_+} \right) \right] t^{1/2},
\end{equation}
and
\begin{equation}
\langle X^2(t) \rangle = 2  \left[ D_+ + \beta(\alpha) (D_- - D_+) \right] t.
\label{eq:MSD}
\end{equation}
For any interpretation, the MSD is linear, i.e., it resembles normal diffusion with an effective diffusion coefficient that depends on $\alpha$. However, except for $\alpha=0$, the process is not centered and the mean scales as $t^{1/2}$. A summary of the results for the It\^o, Stratonovich, and  HK interpretations is shown in Table~\ref{tab:interpretation}.

\begin{table*}
\caption{Summary of the probabilistic analyses for the It\^o ($\alpha = 0$), Stratonovich ($\alpha = 1/2$), and  HK ($\alpha = 1$) interpretations.
\label{tab:interpretation}}
\begin{ruledtabular}
\begin{tabular}{cccc}
 & Mean $\langle X(t) \rangle$ & MSD $\langle X^2(t) \rangle$ & $\beta$ \\ \hline \\

 It\^o &  $0$ &  $ \displaystyle 2 \; \sqrt{D_+ D_-} \; t$ & $\left( \displaystyle  1+ \sqrt{ D_-/D_+} \; \right)^{-1}$  \\ \\ 
 
 Stratonovich  & $\displaystyle \frac{1 }{\sqrt{\pi}} \left( \sqrt{D_+} - \sqrt{D_-} \; \right) t^{1/2}$ & $\displaystyle \left( D_+ + D_- \right) t$ & $1/2$ \\ \\
 
 HK  &  $\displaystyle \frac{2 }{\sqrt{\pi}} \left(\sqrt{D_+} - \sqrt{D_-} \right) t^{1/2}$ & $\displaystyle 2 \left( D_+ + D_- - \sqrt{D_+ D_-} \right) t$ & $\left( \displaystyle  1+ \sqrt{ D_+/D_-} \; \right)^{-1}$  \\ \\
\end{tabular}
\end{ruledtabular}
\end{table*}

To visualize the role of the interpretation, we analyze a process with $D_-=2$ and $D_+=1$, via analytical solutions and numerical simulations. The probability of finding the tracer in the region $x<0$, following Eq.~(\ref{eq:beta}), is presented in Fig.~\ref{fig:PDF}(a). Figures~\ref{fig:PDF}(b) and~\ref{fig:PDF}(c) show the mean $\left<X(t)\right>$ and MSD $\left<X^2(t)\right>$ for the three considered interpretations ($\alpha=0, 1/2, 1$). Next, the PDFs of the position at $t=1024$ are shown in Fig.~\ref{fig:PDF}(d)-(f). In the three cases, the numerical integration of the master equation (Eq.~(\ref{eq:master})) agrees with the analytical expression (Eqs.~(\ref{eq:gen_PDF}), (\ref{eq:beta})).

Often, trajectories of individual particles are analyzed in terms of the time-averaged MSD (TAMSD) \cite{Manzo2015,Munoz2021,Sabri2020}, defined in an observation time $T$, at lag-time $\tau$, as
\begin{equation}
\overline{\delta^2(\tau)} = \frac{1}{T-\tau} \int_{0}^{T-\tau} \left[X(\tau+t) - X(t) \right]^2 dt.
\end{equation}
This quantity is often suitable for the analysis of experimental data, with a limited number of trajectories. 

Brownian motion is an ergodic process where the TAMSD is
$\overline{\delta^2(\tau)} = 2D\tau$. In the two-phase system, the time a tracer spends on one side before switching to the other is the first return time of a standard Brownian motion. The occupation time fraction in the region $x<0$ is a random variable $f_{\alpha}$, and its mean yields the mean of the TAMSD \cite{SupplMat}, 
\begin{equation}
\left<\overline{\delta^2(\tau)}\right> \sim 2\left[D_+ +(D_- -D_+)\beta(\alpha)\right]\tau,
\label{eq:TAMSDmean}
\end{equation}
which shows that $\left<\overline{\delta^2(\tau)}\right> = \left<X^2(\tau)\right>$, in contrast to most non-stationary processes. However, the TAMSD remains a random variable at long realization times, a signature of weak ergodicity breaking.
We consider the coefficient of variation (CV) of the TAMSD, defined as the ratio between the standard deviation $\sigma_{\delta^2}$ and the mean, i.e., $\text{CV} = \sigma_{\delta^2} / \langle \overline{\delta^2(\tau)} \rangle.$

The ergodicity breaking parameter is the square of the CV, which is often used in the study of weak non-ergodicity \cite{Metzler2014}. 
By computing the variance of $f_{\alpha}$ \cite{SupplMat}, we obtain
\begin{equation}
\text{CV} \sim \frac{\displaystyle |D_- - D_+| \sqrt{\beta(\alpha) [1 - \beta(\alpha)]/2}}{D_+ + (D_- - D_+) \beta(\alpha)}.
\label{eq:CV}
\end{equation}
as shown in Supplementary Fig. 1.

A natural question that arises is, given a physical system, how should experimental measurements in the vicinity of an interface guide the choice of the interpretation parameter? The most detailed measurements of particle dynamics are obtained using single-particle tracking. Using this method, diffusion coefficients and the PDF of localization can be obtained on both sides of the two-phase system. For proper initial conditions, trajectories should start when the tracer is at the interface. A continuous density indicates $\alpha=1$  (HK prescription, Fig.~\ref{fig:PDF}(f)). Otherwise, a discontinuity in the PDF at the interface indicates $\alpha\neq 1$, and the interpretation is dictated by the measured probability of being on one side ($\beta$). By inverting Eq.~(\ref{eq:beta}),
\begin{equation}
\alpha = \frac{1}{2} - \frac{\ln (\beta^{-1} - 1)}{\ln(D_-/D_+)}.
\label{eq:alpha_b}
\end{equation}

\begin{figure}
    \includegraphics[width=\columnwidth]{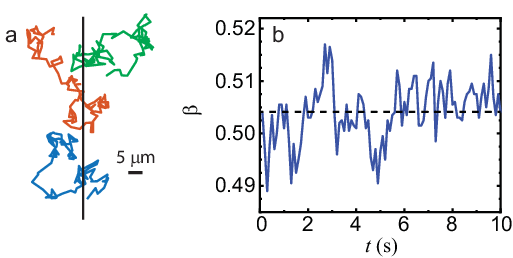}
    \caption{ Inference of the interpretation from a set of trajectories in a two-phase system.
    (a) Representative simulated two-dimensional trajectories of BSA in dextran-rich and PEG-rich phases. The black vertical line delineates the interface. 
    (b) Fraction of trajectories for which the tracer is found on the left yields $\beta(t)$. In this case, the time-average of $\beta(t)$ for $100$ data points is $\overline{\beta}=0.504$, shown as a dashed line. 
    }
    \label{fig:alpha}
\end{figure}

To illustrate the proposed methodology, we employ simulations that represent an experimental system. Our model system is bovine serum albumin (BSA) labeled with AlexaFluor fluorophores within two aqueous phases containing dextran and polyethylene glycol (PEG) \cite{munchow2008protein}. The left phase contains $13.2$ wt\% dextran $500000$, and the right phase contains $7.1$ wt\% PEG $6000$, yielding diffusion coefficients for BSA of $14$ and $24$ $\mu\text{m}^2/\text{s}$, respectively. For these simulations, we assume the system follows the Stratanovich interpretation ($\alpha=1/2$), and data are collected every $0.1$ s. $2000$ trajectories of $100$ data points each are obtained. This combination of trajectory number and length is experimentally realistic in single-particle tracking \cite{Sabri2020}. Figure~\ref{fig:alpha}(a) shows  representative simulations of BSA trajectories. 

While it is possible to use the mean, MSD, or TAMSD CV to infer the interpretation $\alpha$, we showcase a simplified method using only the fraction of trajectories to the left of the interface at a given time. If the selected interpretation is correct, the predictions for MSD and CV should ensue. Figure~\ref{fig:alpha}(a) presents $\beta(t)$ obtained as the fraction of points to the left of the interface. The time average for this dataset is $\overline{\beta}=0.504$, which yields an interpretation $\alpha=0.470$, according to Eq.~(\ref{eq:alpha_b}). To assess estimation error, we repeated the described procedure $100$ times. The statistical results are $\overline{\beta}=0.501\pm 0.007$ and, in turn, $\alpha=0.49 \pm 0.06$ (mean $\pm$ standard deviation). Histograms of $\overline{\beta}$ and $\alpha$ are presented in Supplementary Fig. 2. 

The simplicity of a system with a single interface allows us to obtain analytical results and highlights the differences between interpretations. 
Even though our approach is for the general case, we focus our analysis on the three most widely used interpretations. Each of these prescriptions has appealing strengths: with the It\^o interpretation, the mean is always zero, a consequence of this process being a martingale, that is, the conditional expectation of the next value in a sequence equals the current value, regardless of its history \cite{Oksendal2013}. While
the It\^o interpretation maintains a constant mean position (martingale property), the
Stratonovich one preserves the median in the two-phase system. With the Stratonovich interpretation, a tracer has equal probabilities of being on either side of the origin, i.e., $\probP[X(t)<0] = 1/2$.  
With the HK interpretation, the Fokker-Planck equation coincides with that obtained from combining Fick's first law of diffusion and the continuity equation, and the PDF of displacements is continuous at the origin. The Fokker-Planck equations for the three interpretations are listed in Supplementary Table I. 

To make sense of the results for different interpretations in inhomogeneous systems, we briefly consider a system in a box with reflecting boundary conditions. As this system equilibrates, the distribution converges to $p(x)\sim D^{\alpha-1}(x)$ \cite{SupplMat}. From this Boltzmann distribution,  
we obtain an effective potential $U(x) = -k_B T \ln p(x)$ of the form 
\begin{equation}
 U(x) = k_B T (1- \alpha) \ln D(x) + \mathrm{const},
\end{equation}
which has a discontinuity, except for the HK case. 
Overall, the changes in $D$ imply changes in the interactions between the tracer and its surroundings.
The properties of the non-equilibrium PDFs follow from the  discontinuity in the potential, i.e. 
from an unbalanced force in the vicinity of the boundary towards the domain with a lower diffusion coefficient.

The process treated in the main part of this work (Fig.~\ref{fig:PDF}) is not confined and, as such, it describes an out-of-equilibrium system that does not reach a steady state. Due to the absence of
confinement, a so-called infinite density emerges \cite{aaronson1997introduction,aghion2020infinite}. Several statistical properties of these processes and their relation
with infinite ergodic theory have been previously analyzed \cite{Aghion2019,Leibovich2019}. 
Despite these complexities, the probability $\beta$ does not depend on time, facilitating the analysis of experimental results. Given the interpretation, using infinite ergodic theory formalism, it is possible to obtain thermodynamical properties of the system
and relations between time- and ensemble-averaged observables.

One interesting aspect is that Brownian motion with alternating diffusivities maintains the linear time behavior of the MSD, regardless of the interpretation. However, the effective diffusion coefficient depends on the interpretation. With the It\^o interpretation, the effective diffusivity is equal to the geometric mean of the two diffusion coefficients, while with the Stratonovich interpretation, the effective diffusivity is equal to their arithmetic mean. 

Throughout the study of a simple two-phase heterogeneous environment, we have shown how the overall characterization of the motion depends on the interpretation parameter. This result encourages researchers modeling diffusion in heterogeneous environments to consider which interpretation suits best the physics of their system. These results do not imply one interpretation is better than the others, or that choosing one makes the others incorrect. Selecting an interpretation simply states how the Langevin equation will be treated. 

\begin{acknowledgments}
This work was supported by the National Science Foundation Grant 2102832 (to DK) and the National Science Centre, Poland, projects 2020/37/B/HS4/00120 (to AW) and 2023/07/X/ST1/01139 (to MB).
\end{acknowledgments}


\providecommand{\noopsort}[1]{}\providecommand{\singleletter}[1]{#1}%

\end{document}


\preprint{APS/123-QED}

\title{Langevin equation in heterogeneous landscapes: how to choose the interpretation \\ \vspace{0.3 cm}
Supplemental Material}

\author{Adrian Pacheco-Pozo}
    \affiliation{Department of Electrical and Computer Engineering and School of Biomedical Engineering, Colorado State University, Fort Collins, CO 80523, USA}
 
\author{Micha{\l} Balcerek}%
    \affiliation{Faculty of Pure and Applied Mathematics, Hugo Steinhaus Centre, Wroc{\l}aw University of Science and Technology, 50-370 Wrocław, Poland}%

\author{Agnieszka Wy{\l}omanska}
    \affiliation{Faculty of Pure and Applied Mathematics, Hugo Steinhaus Centre, Wroc{\l}aw University of Science and Technology, 50-370 Wrocław, Poland}%

\author{Krzysztof Burnecki}
    \affiliation{Faculty of Pure and Applied Mathematics, Hugo Steinhaus Centre, Wroc{\l}aw University of Science and Technology, 50-370 Wrocław, Poland}%

\author{Igor M. Sokolov}
\affiliation{Institut f\"{u}r Physik and IRIS Adlershof, Humboldt University Berlin, Newtonstrasse 15, 12489 Berlin, Germany}

\author{Diego Krapf}
    \affiliation{Department of Electrical and Computer Engineering and School of Biomedical Engineering, Colorado State University, Fort Collins, CO 80523, USA}


\maketitle

This Supplemental Material consists of mathematical details (I to VI), methods (VII), Supplementary Figures 1 and 2, and Supplementary Table I. 
\begin{enumerate}[I.]
 \item Generalized two-piece Gaussian distribution.
 \item Verification that the identified PDF is the solution to the Fokker-Planck equations.
 \item General form of $\beta$.
 \item Inhomogeneous system in a box.
 \item Time-averaged MSD (TAMSD).
 \item Distribution of TAMSD.
 \item Procedure to generate synthetic trajectories.
\end{enumerate}

{\bf Supplementary Table I.} Fokker-Planck equation.

{\bf Supplementary Figure 1.} Coefficient of variation of the TAMSD.

{\bf Supplementary Figure 2.} Inference of the interpretation from a set of trajectories in a two-phase system.

\section{Generalized two-piece Gaussian distribution}
Let us introduce a \textit{generalized} two-piece Gaussian distribution. We start by recalling the standard two-piece Gaussian (normal) distribution \cite{Wallis2014}, 
\begin{equation}
f(x)= \left\{ 
\begin{array}{lcc} 
\displaystyle A \exp \left( - \frac{x^2}{2 \sigma^2_-} \right) & \text{if} & x < 0 \\ \\
\displaystyle A \exp \left( - \frac{x^2}{2 \sigma^2_+} \right)  & \text{if} & x \geq 0
\end{array} 
\right.
\end{equation}
\\
with normalization constant
\begin{equation}
A = \left[ \sqrt{2 \pi} \; \frac{\sigma_- + \sigma_+}{2} \right]^{-1}.
\end{equation}
This distribution can be rewritten in the form
\begin{equation}
f(x)= \left\{ 
\begin{array}{lcc} 
\displaystyle \frac{2 \beta}{ \sqrt{2 \pi \sigma_-^2}} \exp \left( - \frac{x^2}{2 \sigma^2_-} \right) & \text{if} & x < 0, \\ \\
\displaystyle \frac{2 (1 - \beta)}{ \sqrt{2 \pi \sigma_+^2}}  \exp \left( - \frac{x^2}{2 \sigma^2_+} \right)  & \text{if} & x \geq 0,
\end{array} 
\right.
\label{eq:gen_two_gaus}
\end{equation}
with $\beta$ being the probability that the random variable $X$ is less than zero,
\begin{equation}
\beta = \int_{-\infty}^0 f(x) dx = \left( 1 + \frac{\sigma_+}{\sigma_-} \right)^{-1}.
\end{equation}

We generalize the two-piece Gaussian distribution (Eq.~(\ref{eq:gen_two_gaus})) for an arbitrary parameter $\beta$ and proceed to study the properties of this generalized distribution. The mean is found as 
\begin{align}
\langle X \rangle &= \beta \int_{-\infty}^{0} \frac{2x}{\sqrt{2 \pi \sigma_-^2 }} \exp \left(- \frac{x^2}{2 \sigma_-^2} \right)  dx \; + \nonumber \\
&+ (1-\beta) \int_{0}^{\infty} \frac{2x}{\sqrt{2 \pi \sigma_+^2 }} \exp \left(- \frac{x^2}{2 \sigma_+^2} \right) dx \\
& = \sqrt{\frac{2 }{\pi}} \sigma_+ \left[ 1 - \beta \left( 1 + \frac{\sigma_-}{\sigma_+} \right) \right].
\end{align}
The distribution has a zero mean only when $\beta=[1 + (\sigma_-/\sigma_+)]^{-1}$. The second moment is
\begin{align}
\langle X^2 \rangle &= \beta \int_{-\infty}^{0} \frac{2x^2}{\sqrt{2 \pi \sigma_-^2 }} \exp \left(- \frac{x^2}{2 \sigma_-^2} \right)  dx \; + \nonumber \\
&+ (1-\beta) \int_{0}^{\infty} \frac{2x^2}{\sqrt{2 \pi \sigma_+^2 }} \exp \left(- \frac{x^2}{2 \sigma_+^2} \right) dx \\
&= \sigma_+^2 +\beta \left( \sigma_-^2 - \sigma_+^2 \right).
\end{align}

\section{Verification that the identified PDF is the solution to the Fokker-Planck equation}

We can rewrite the Fokker-Planck equation (Eq.~(4) of the main text) in the form
\begin{equation}
\frac{\partial p(x,t)}{\partial t} = \frac{\partial}{\partial x} \left[ (1 - \alpha) \frac{\partial D(x)}{\partial x} + D(x) \frac{\partial}{\partial x} \right] p(x,t).
\label{eq:Fokker_Planck}
\end{equation}
To verify that the identified PDF (Eq.~(8) in the main text) is the solution to the Fokker-Planck equation, we substitute it into Eq.~(\ref{eq:Fokker_Planck}). For this purpose, we rewrite the PDF as
\begin{equation}
p(x,t;\alpha) = \frac{2 \beta(x,\alpha)}{\sqrt{4 \pi D(x) t}} \exp \left( - \frac{x^2}{4 D(x) t} \right),
\label{eq:gen_sol}
\end{equation}
with
\begin{equation}
D(x) = (D_+ - D_-) U(x) + D_-,
\label{eq:DFunction}
\end{equation}
and
\begin{equation}
\beta(x,\alpha) = [1 - 2 \beta(\alpha) ] U(x) + \beta(\alpha),
\end{equation}
where $U(x)$ is the Heaviside step function. Since the step function has a discontinuity at zero, and, therefore, its derivatives do not exist at this point, we consider the approximation
\begin{equation}
\mathcal{U}(x) = \frac{1}{1 + \exp(-kx)}
\end{equation}
which, in the limit $k \to \infty$, converges to the step function ${U}(x)$. With this approximation, we rewrite the diffusion landscape
\begin{equation}
\mathcal{D}(x) = \frac{D_+ - D_- }{1 + \exp(-kx)} + D_-,
\end{equation}
which, in the limit $k \to \infty$, converges to $D(x)$. We also define the function 
\begin{equation}
\mathcal{B}(x,\alpha) = \frac{1 - 2 \beta(\alpha) }{1 + \exp(-kx)} + \beta(\alpha),
\end{equation}
which converges to $\beta(x,\alpha)$. Finally, we consider the analytical approximation of the PDF
\begin{equation}
\mathcal{P}(x,t;\alpha) = \frac{2 \mathcal{B}(x,\alpha)}{\sqrt{4 \pi \mathcal{D}(x) t}} \exp \left( - \frac{x^2}{4 \mathcal{D}(x) t} \right),
\label{eq:analilt_PDF}
\end{equation}
which converges to $p(x,t;\alpha)$ as $k \to \infty$.
For convenience, let us express the function $\mathcal{B}(x,\alpha)$ in terms of $\mathcal{D}(x)$ as
\begin{equation}
\mathcal{B}(x,\alpha) = A_1(\alpha) \mathcal{D}(x) + A_2(\alpha),
\end{equation}
with
\begin{equation}
A_1(\alpha) = \frac{1 - 2 \beta(\alpha)}{D_+ - D_-},
\end{equation}
and
\begin{equation}
A_2(\alpha) = \beta(\alpha) \frac{D_+ + D_-}{D_+ - D_-} - \frac{D_-}{D_+-D_-}.
\end{equation}
Additionally, to simplify the notation, we drop all the $\alpha$ dependencies.

Now, let us first calculate the time-derivative of Eq.~(\ref{eq:analilt_PDF})
\begin{equation*}
\frac{\partial}{\partial t} \mathcal{P}(x,t) = 
 \frac{\mathcal{B}(x) \left[ x^2 - 2 \mathcal{D}(x) t \right] }{ 4 \sqrt{\pi } \mathcal{D}(x)^{3/2} t^{5/2}}  \exp \left( - \frac{x^2}{4 \mathcal{D}(x) t} \right).
\end{equation*}
Then, we take the limit $k \to \infty$, so as to find the l.h.s. of the Fokker-Planck equation~(\ref{eq:Fokker_Planck})
\begin{equation}
\frac{\partial}{\partial t} p(x,t) = \frac{\beta(x) \left[ x^2 - 2 D(x) t \right]  }{ 4 \sqrt{\pi } D(x)^{3/2} t^{5/2}} \exp \left( - \frac{x^2}{4 D(x) t} \right). 
\label{eq:lhs}
\end{equation}
To find the r.h.s of the Fokker-Planck equation, we calculate the terms involving the space derivatives,
\begin{widetext}
\begin{equation}
(1- \alpha) \frac{\partial \mathcal{D}(x)}{\partial x} \mathcal{P}(x,t) = \frac{(1 - \alpha)}{\sqrt{\pi t}} [A_1 \mathcal{D}(x) + A_2 ] \exp \left[ - \frac{x^2}{ 4 \mathcal{D}(x) t} \right]   \mathcal{D}(x)^{-1/2} \mathcal{D}^{\prime}(x),
\end{equation}
\begin{multline}
\mathcal{D}(x) \frac{\partial}{\partial x} \mathcal{P}(x,t) = \\
=\frac{1}{2 \sqrt{\pi t}} \exp \left[ - \frac{x^2}{ 4 \mathcal{D}(x) t} \right] \left[ A_1 \mathcal{D}^{1/2} \mathcal{D}^{\prime}(x) - \frac{1}{2} \mathcal{B}(x) \mathcal{D}(x)^{-1/2} \mathcal{D}^{\prime}(x) - \frac{x}{2 t} \mathcal{B}(x) \mathcal{D}(x)^{-1/2} + \frac{x^2}{4t} \mathcal{B}(x) \mathcal{D}(x)^{-3/2} \mathcal{D}^{\prime}(x)  \right],
\end{multline}
with $\mathcal{D}^{\prime}(x) = \partial \mathcal{D}(x) / \partial x$.
Then,
\begin{align}
& \frac{\partial}{\partial x} \left[ (1 - \alpha) \frac{\partial \mathcal{D}(x)}{\partial x} + \mathcal{D}(x) \frac{\partial}{\partial x} \right] \mathcal{P}(x,t)  = \nonumber  \\
& \nonumber \frac{1}{16 \sqrt{\pi} \mathcal{D}(x)^{7/2} t^{5/2} } \exp \left[-\frac{x^2}{4 \mathcal{D}(x) t} \right] \{ x^2 \mathcal{D}(x) \mathcal{D}'(x) [ \mathcal{D}'(x) ( A_1 x^2 - 4 ( \alpha + 1 ) A_2 t ) - 4 A_2 x ] +  \\
& \nonumber + 4 \mathcal{D}(x)^3 [ t \mathcal{D}''(x) ( (2 - 4 \alpha) A_2 t + A_1 x^2 ) + (3-2 \alpha ) A_1 t^2 \mathcal{D}'(x)^2 + 2 ( \alpha -1 ) A_1 t x \mathcal{D}'(x) + A_1 x^2 - 2 A_2 t ] +  \\
& \nonumber +4 \mathcal{D}(x)^2 [\mathcal{D}'(x) \left(2 ( \alpha + 1 ) A_2 t
   x-A_1 x^3\right)+t \mathcal{D}'(x)^2 \left((2 \alpha -1) A_2 t-(\alpha-1) A_1 x^2\right)+A_2 x^2 (t
   \mathcal{D}''(x)+1)] \\
&  -8 A_1 t \mathcal{D}(x)^4 \left((2 \alpha-3) t \mathcal{D}''(x)+1\right)+A_2 x^4 \mathcal{D}'(x)^2 \},
\label{eq:rhs}
\end{align}
\end{widetext}
with $\mathcal{D}^{\prime\prime}(x) = \partial^2 \mathcal{D}(x) / \partial x^2$.

By taking the limit $k \to \infty$ and using the fact that 
\begin{equation}
\lim_{k \to \infty} \mathcal{D}(x)^{\prime} = 0,
\end{equation}
and
\begin{equation}
\lim_{k \to \infty} \mathcal{D}(x)^{\prime\prime} = 0,
\end{equation}
for $x\neq0$, all terms in Eq.~(\ref{eq:rhs}) having either a first or a second derivative of $\mathcal{D}(x)$ vanish. Since we are dealing with a continuous variable, a single point has zero measure and therefore does not have any effect when computing probabilistic quantities. 

Finally, one has 
\begin{align*}
\text{l.h.s.} = \frac{\beta(x) \left[ x^2 - 2 D(x) t \right]  }{ 4 \sqrt{\pi } D(x)^{3/2} t^{5/2}} \exp \left[ - \frac{x^2}{4 D(x) t} \right]. 
\end{align*}
Since the l.h.s. (Eq.~(\ref{eq:lhs})) is equal to the r.h.s. (given by the latter equation), it verifies that Eq.~(\ref{eq:gen_sol}) is the solution to the Fokker-Planck equation~(\ref{eq:Fokker_Planck}).

\section{General form of $\beta$}

Let us start from the Fokker-Planck equation in the form
\begin{equation}
\frac{\partial p(x,t)}{\partial t} = \frac{\partial}{\partial x}  \left\{ D^{\alpha}(x) \frac{\partial}{\partial x} \left[ D^{1-\alpha}(x) p(x,t) \right] \right\},
\label{eq:Fokker_Planck1}
\end{equation}
and note that the temporal derivative has to be well-defined at any time and for any position.
The Fokker-Planck equation is a combination of the continuity equation $\frac{\partial p(x,t)}{\partial t} = - \frac{\partial}{\partial x} j(x,t)$,
where $j(x,t)$ is the diffusive flux, and a generalized Fick's first law
\begin{equation}
 j(x,t) = - D^{\alpha}(x) \frac{\partial}{\partial x} \left[ D^{1-\alpha}(x) p(x,t) \right].
 \label{eq:flux}
\end{equation}
This means that the expression in curvy brackets in Eq.~(\ref{eq:Fokker_Planck1}), the flux, is differentiable
(i.e., at least continuous) everywhere. Since $D(x)$ nowhere vanishes, we can write
\begin{equation}
\frac{\partial}{\partial x} \left[ D^{1-\alpha}(x) p(x,t) \right] =  - \frac{1}{D^{\alpha}(x)} j(x,t).
\label{eq:Match}
\end{equation}
Using the expression for the diffusivity landscape $D(x)$ defined in Eq.~(\ref{eq:DFunction}) and integrating both sides of Eq.~(\ref{eq:Match}) from $-\epsilon$ to $\epsilon$ in the vicinity of the origin, we obtain
\begin{equation}
D^{1-\alpha}(\epsilon) p(\epsilon,t) - D^{1-\alpha}(-\epsilon) p(-\epsilon,t) = - \int_{-\epsilon}^\epsilon \frac{1}{D^{\alpha}(x)} j(x,t).
\end{equation}
Since $j(x,t)$ is continuous, the mean value theorem states that there exists a point $x^* \in (-\epsilon, \epsilon)$ such that $j(x^*,t)$ can be taken out of the integral, 
\begin{multline*}
D^{1-\alpha}(\epsilon) p(\epsilon,t) - D^{1-\alpha}(-\epsilon) p(-\epsilon,t) =\\
=- j(x^*,t) \left(\frac{\epsilon}{D_-^{\alpha}} + \frac{\epsilon}{D_+^{\alpha}} \right).
\end{multline*}
The r.h.s. vanishes for $\epsilon \to 0$, thus 
\begin{equation}
D_+^{1-\alpha} p(0_+,t) - D_-^{1-\alpha} p(0_-,t) = 0,
\label{eq:Dstep}
\end{equation}
and the PDF has to show a finite jump for $\alpha \neq 1$. 

We can now use this matching condition to build the solution from the partial solutions on the left and right half-lines (Eq.~(8) in the main text),
\begin{equation}
 p(0_+,t) = \frac{2(1-\beta)}{\sqrt{4 \pi D_+ t}}, \qquad p(0_-,t) = \frac{2\beta}{\sqrt{4 \pi D_- t}},
\end{equation}
which can be substituted into Eq.~(\ref{eq:Dstep})
\begin{equation}
 (1-\beta) D_+^{(1/2-\alpha)} = \beta D_-^{(1/2-\alpha)} 
\end{equation}
yielding the solution
\begin{equation}
 \beta(\alpha) = \left[ 1 + \left( \frac{D_-}{D_+} \right)^{(1/2-\alpha)} \right]^{-1}.
\end{equation}

\section{Inhomogeneous system in a box}

While in the main text we discuss the natural boundary condition corresponding to the PDFs vanishing at infinity, 
we can also consider a system in a box with reflecting boundary conditions at two impenetrable walls posed at $x=-L$ and at $x=L$. In this case, in the course of time an equilibrium distribution $p(x)$ characterized by 
vanishing flux establishes itself: from Eq.~(\ref{eq:flux}), under the requirement $j(x)=0$, we get
\begin{equation}
 D^{1-\alpha}(x) p(x) = const,
\end{equation}
where the corresponding constant arises from the normalization. Furthermore, Eq.~(\ref{eq:Dstep}) shows that this condition holds also in the non-equilibrium case. The normalization yields 
\begin{equation}
p(x) = \frac{1}{L(D_+^{\alpha}+ D_-^{\alpha})} D^{\alpha-1}(x),
\end{equation}
with $D(x) = D_-$ for $x<0$ and $D_+$ for $x\geq 0$. Calculating the potential of the mean force $U(x) = - k_B T \ln p(x)$, i.e. comparing our equilibrium PDF with the Boltzmann distribution
\begin{equation}
p(x) \propto \exp[-U(x)/k_B T],    
\end{equation}
we see that the potential $U(x)$ of the mean force is 
\begin{equation}
 U(x) = k_B T (1- \alpha) \ln D(x) + \mathrm{const},
\end{equation}
which exhibits a jump on the boundary of the media, except for the H\"anggi-Klimontovich case.

\section{Time-averaged MSD (TAMSD)}

In order to analyze the trajectories that emerge from the master equation~(Eq.~(5) of the main text), we consider the equivalence of the master equation to a continuous-time random walk (CTRW) with exponential waiting time distribution \cite{Klafter2011}. This is seen by expressing the transition rates $\omega_{i \to j}$ as the ratio between the probability of transitioning from $i$ to $j$, denoted as $\probP(i \to j)$, and the mean waiting time at site $i$ (the current particle position), denoted by $\tau_i$. Namely, $\omega_{i \to j} = \probP(i \to j) / \tau_i$. Then, using Eq.~(6) of the main text one obtains 
\begin{equation}
\probP(i \to j) = \frac{D_k^{\alpha}}{D_{i+1/2}^{\alpha} + D_{i-1/2}^{\alpha}},
\label{eq:trans_prob}
\end{equation}
with $k = (i + j)/2$, and
\begin{equation}
\tau_i = \frac{a^2}{D_i^{1-\alpha} (D_{i+1/2}^{\alpha} + D_{i-1/2}^{\alpha})},
\label{eq:mean_wt}
\end{equation}
where we considered that the particle always performs a jump, i.e., $\sum_j \probP(i \to j) = 1$. The above process allows us to rewrite the master equation in the form
\begin{equation}
\frac{d p_i(t)}{dt} = \sum_{m = 1}^{2} \frac{\probP(j_m \to i)}{\tau_{j_m}} p_{j_m}(t) - \frac{1}{\tau_{i}} p_{i}(t), 
\end{equation}
with $j_1 = i-1$ and $j_2 = i+1$. From the latter, it is clear how the master equation is related to a CTRW scheme: a particle at lattice point $i$ waits a time $\delta t$, taken from an exponential distribution with mean $\tau_i$,
before making an instant jump to either the left ($i - 1$) with probability ${\probP(i \to i-1)}$ or to the right ($i+1$) with probability ${\probP(i \to i+1)}$. The final result is a trajectory $X(t_i)$ with $t_i = \sum_{j=1}^{i} \delta t_j$. One can then evenly sample such trajectory so as to estimate the TAMSD.

For the different interpretations, the waiting times are counterintuitive. For the It\^o interpretation ($\alpha = 0$), the transition probability is symmetric, i.e. 
 \begin{equation}
\probP(i \to j) = \frac{1}{2},
\end{equation}
corresponding to a unbiased random walk. This goes hand-in-hand with the fact that the system under It\^o interpretation is a centered process. The mean waiting time at a site $i$ takes the form
\begin{equation}
\tau_i = \frac{a^2}{2 D_i},
\end{equation}
which only depends on the diffusivity at the current location of the particle. 
Furthermore, from Eqs.~(\ref{eq:trans_prob}) and~(\ref{eq:mean_wt}) it can be seen that only for $\alpha = 0$, i.e. It\^o interpretation, one has an unbiased random walk with a mean waiting time that only depends on the diffusivity at the site where the particle is currently located. Any other $\alpha$ results in a biased random walk with a non-trivial mean waiting time. 

For the Stratonovich interpretation with $\alpha = 1/2$, the transition probability takes the form
\begin{equation}
\probP(i \to j) = \frac{\sqrt{D_k}}{ \sqrt{D_{i+1/2}} + \sqrt{D_{i-1/2}}},
\end{equation}
and the mean waiting time at the site is
\begin{equation}
\tau_i = \frac{a^2}{\sqrt{D_i} ( \sqrt{D_{i+1/2}} + \sqrt{D_{i-1/2}})}.
\end{equation}
On the other hand, for the HK interpretation with $\alpha = 1$, the transition probability takes the form 
 \begin{equation}
\probP(i \to j) = \frac{D_k}{D_{i+1/2} + D_{i-1/2}},
\end{equation}
and the mean waiting time at the site is
\begin{equation}
\tau_i = \frac{a^2}{D_{i+1/2} + D_{i-1/2}}.
\end{equation} 

\section{Distribution of TAMSD}

In the two-phase system, an individual trajectory alternates between regions of different diffusivities with occupation time fraction in the region $x<0$ being the random variable $f_{\alpha}$. The TAMSD in the long realization time is then \cite{Balcerek2023} 
\begin{equation}
\overline{\delta^2}(\tau)\sim 2 D_+ \tau ( 1 - f_{\alpha} ) + 2 D_- \tau f_{\alpha}. 
\label{Eq:conditionalTAMSD1}
\end{equation}
The occupation times in each region have a heavy-tailed distribution $p_{T\pm}(t)\sim A_{\alpha}^{\pm} t^{-3/2}$  
where $A_{\alpha}^{\pm}$ are the amplitudes for the times spent on the right or left sides, $A_{\alpha}^-=\beta(\alpha)$ and $A_{\alpha}^+=1-\beta(\alpha)$ \cite{Krug1997,Molchan1999,Sanders2012}.
As a consequence, the mean and variance of $f_{\alpha}$ are \cite{Rebenshtok2013}
\begin{equation}
\left<f_{\alpha}\right>=\beta(\alpha)
\end{equation}
 and 
 \begin{equation}
 \sigma^2_{f_{\alpha}}=\beta(\alpha)[1-\beta(\alpha)]/2.
 \end{equation}
Using these expressions, the mean of the TAMSD converges in the long realization time limit to
\begin{equation}
\left<\overline{\delta^2}(\tau)\right> \sim 2\left[D_+ +(D_- -D_+)\beta(\alpha)\right]\tau,
\label{eq:TAMSDmean}
\end{equation}
and the variance of the TAMSD is
\begin{equation}
\sigma^2_{\delta^2} \sim 2\left(D_- -D_+\right)^2\beta(1-\beta)\tau.
\end{equation}
The coefficient of variation is then found,
\begin{equation}
\text{CV} \sim \frac{\displaystyle |D_- - D_+| \sqrt{\beta(\alpha) [1 - \beta(\alpha)]/2}}{D_+ + (D_- - D_+) \beta(\alpha)}.
\label{eq:CV}
\end{equation}
The occupation time fraction $f_\alpha$ follows a Lamperti PDF \cite{Rebenshtok2013,Lejay2018} 
\begin{multline}
p_{F\pm}(f_{\alpha})= \frac{\beta(\alpha)}{\pi [1-\beta(\alpha)]}  \times \\
\times \frac{1}{\sqrt{(1-f_{\alpha})f_{\alpha}} \left\{ \displaystyle f_{\alpha} + \left[ \frac{\beta(\alpha)}{1-\beta(\alpha)} \right]^2 (1-f_{\alpha}) \right\} }.
\label{Eq.Lampertibm}
\end{multline}
Thus, we can directly obtain the PDF of $\overline{\delta^2(\tau)}$ as
\begin{equation}
p_\delta\left(\overline{\delta^2}\right) = \frac{1}{2\left|D_- -D_+\right|\tau} p_{F\pm}\left(\frac{\overline{\delta^2}-2D_+\tau}{2\left(D_- -D_+\right)\tau}\right).
\label{Eq:PDF-TAMSD}
\end{equation}

\section{Procedure to generate synthetic trajectories}

The proposed methodology used to find the interpretation $\alpha$ was tested with synthetic trajectories. Here, we present the Heun algorithm \cite{Ruemelin1982}, which was used to generate these trajectories. The position of the particle $X(t)$
obeys the Langevin equation
\begin{equation}
\frac{d X(t)}{dt} = \sqrt{2 D[X(t)]} \; \xi(t),
\end{equation}
where $\xi(t)$ is a zero-mean, delta-correlated Gaussian white noise. The numerical algorithm used to find a solution for this equation for the Stratonovich interpretation ($\alpha = 1/2$) is the stochastic Heun algorithm. This algorithm is a predictor-corrector method that consists of two steps.

{\bf First}, one predicts the value $X(t_{N})$, denoted as $\widehat{X_N}$, via an Euler-Mayurama procedure
\begin{equation}
\widehat{X_N} = X_{N-1} + \sqrt{2 D\left(X_{N-1}\right)} \cdot \xi_N,
\end{equation}
where $X_0=0$ and $\xi_N$ are normal i.i.d. random variables with variance $\Delta t$, the sampling time.

{\bf Second}, one corrects the predicted value by 
\begin{multline}
X_N = X_{N-1} + \frac{1}{2}\left[\sqrt{2 D(\widehat{X_N})} + \sqrt{2 D(X_{N-1})} \right] \xi_N. 
\end{multline}

\makeatletter
\@fpsep\textheight
\makeatother

\begin{table*}[p!]
\caption{Fokker-Planck equations for the It\^o ($\alpha = 0$), Stratonovich ($\alpha = 1/2$), and  HK ($\alpha = 1$) interpretations.
\label{tab:FP}}
\begin{ruledtabular}
\begin{tabular}{cccc}

 &It\^o &  $\displaystyle \frac{\partial p_I(x,t)}{\partial t} = \frac{\partial^2}{\partial^2 x} \left[ D(x) p_I(x,t) \right]$ &  \\ \\ 
 
 &Stratonovich  & $\displaystyle \frac{\partial p_S(x,t)}{\partial t} = \frac{\partial}{\partial x}  \left\{ \sqrt{D(x)} \frac{\partial}{\partial x} \left[ \sqrt{D(x)} p_S(x,t) \right] \right\}$ & \\ \\
 
 &H\"anggy-Klimontovich  &  $\displaystyle \frac{\partial p_{HK}(x,t)}{\partial t} = \frac{\partial}{\partial x}  \left[ D(x) \frac{\partial  p_{HK}(x,t)}{\partial x} \right]$ & \\ \\

\end{tabular}
\end{ruledtabular}
\end{table*}

\begin{figure*}[p!]
    \centering
    \includegraphics[width=1.2\columnwidth]{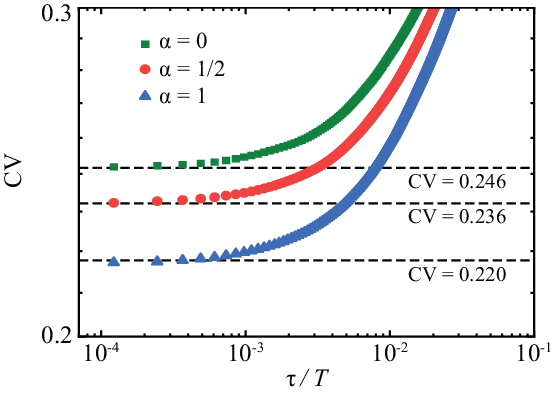}
    \caption{Coefficient of variation (CV) of the TAMSD with $D_-=2$ and $D_+=1$, as a function of the ratio between lag time $\tau$ and realization time $T$, which in this case is $T = 8192$, for three interpretation parameters $\alpha = 0, 1/2,$ and $1$, corresponding respectively to the It\^o, Stratonovich and HK interpretations. The dashed lines correspond to the  CV in the long realization time limit ($T \to \infty$), and is given by Eq.~(15).}
    \label{fig:CV}
\end{figure*}

\begin{figure*}[p!]
    \includegraphics[width=1.4\columnwidth]{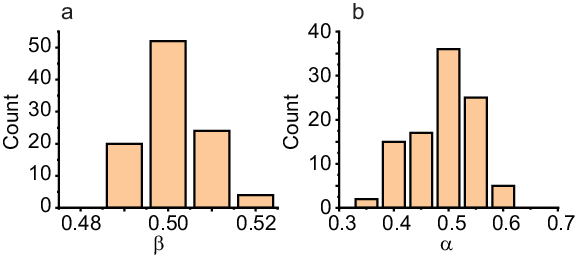}
    \caption{ Inference of the interpretation from a set of trajectories in a two-phase system. (a) Histogram of $\overline{\beta}$ as inferred from $100$ repetitions of the procedure with $2000$ realizations each. 
    (b) Histogram of $\alpha$ obtained from $\overline{\beta}$.
    }
\end{figure*}

\clearpage

\providecommand{\noopsort}[1]{}\providecommand{\singleletter}[1]{#1}%